\documentclass[smallcondensed]{svjour3}
\smartqed
\usepackage{graphicx}

\begin{document}

\title{Tidal evolution of close-in exoplanets in co-orbital configurations}

\titlerunning{Tidal evolution of exo-Trojans} 

\author{Adri\'an Rodr\'iguez \and Cristian A. Giuppone \and Tatiana A. Michtchenko}

\authorrunning{A. Rodr\'iguez et al.} 

\institute{A. Rodr\'iguez \and T. A. Michtchenko \at
Instituto de Astronomia, Geof\'{\i}sica e Ci\^encias Atmosf\'ericas, IAG\,-\,USP,\\
Universidade de S\~ao Paulo, S\~ao Paulo, SP, Brazil\\
\email{adrian@astro.iag.usp.br} \and  C.A. Giuppone \at Observatorio Astron\'omico, Universidad Nacional de C\'ordoba, IATE, C\'ordoba, Argentina\\}

\maketitle

\begin{abstract}


In this paper, we study the behavior of a pair of co-orbital planets, both orbiting a central star on the same plane and undergoing tidal interactions. Our goal is to investigate final orbital configurations of the planets, initially involved in the 1/1 mean-motion resonance (MMR), after long-lasting tidal evolution. The study is done in the form of purely numerical simulations of the exact equations of motions accounting for gravitational and tidal forces. The results obtained  show that, at least for equal mass planets, the combined effects of the resonant and tidal interactions provoke the orbital instability of the system, often resulting in collision between the planets. We first discuss the case of two hot-super-Earth planets, whose orbital dynamics can be easily understood in the frame of our semi-analytical model of the 1/1 MMR. Systems consisting of two hot-Saturn planets are also briefly discussed.


\end{abstract}

\section{Introduction}

In contrast with strong mean-motion resonances (MMR), such as 2/1 and 3/2 MMR (see Beaug\'e et al. 2012) there are no known pairs in co-orbital configuration, despite the continuously increasing number of discovered exoplanets. This intriguing fact can be investigated in the context of a past large-scale planetary migration and capture due to, for instance, interactions with the gaseous protoplanetary disk. Several theories about the formation, dissipative evolution and possible detection of exo-Trojans have been developed and will be discussed later in this section.

The first detailed analysis of hypothetical co-orbital exoplanets was done in Laughlin \& Chambers (2002), focusing the difficulties in fitting of the RV data of these systems. More recently, Hadjidemetriou et al. (2009) studied the topology of the phase space and the long term evolution of two-planet systems in the vicinity of the exact 1/1 MMR in the conservative case and for several planetary mass ratios. Hadjidemetriou \& Voyatzis (2011) included a drag force and simulated the dynamical evolution of a two-planet system initially trapped in a stable 1/1 resonant periodic motion. Under this drag force the system migrates along the families of periodic orbits and is finally trapped in a satellite orbit. Giuppone et al. (2010) analyzed the stability regions and families of periodic orbits of two planets locked in a co-orbital configuration using a semi-analytical method. They found two new asymmetric solutions which do not exist in the restricted three-body problem. These solutions were also obtained by Robutel \& Pousse (2013), who developed an analytical Hamiltonian formalism adapted to the study of the motion of two planets in co-orbital resonance for near coplanar and near circular orbits.

Another kind of dissipative evolution in planetary systems is due to tidal interactions of short-period planets with the central star, which originate changes in orbital elements and rotational periods of the planets. For a planet orbiting a slow rotating star, the tidal effects lead to orbital decay, orbital circularization and rotation synchronization on timescales which depend on physical parameters and initial orbital configurations of interacting bodies (see Dobbs-Dixon et al. 2004; Ferraz-Mello et al. 2008; Rodr\'iguez et al. 2011a, Michtchenko \&  Rodr\'iguez 2011 and references therein). Some recent studies address the problem of tidal interaction in two-planet resonant systems (see Papaloizou 2011; Delisle et al. 2012). They show that, as tides tend to circularize the planetary orbits, the planets repel each other, in such a way that the period ratios at the end of the evolution are slightly larger than the corresponding nominal low-order resonant values. This feature seems to be a natural outcome of slow dissipative evolution (Lithwick \& Wu 2012; Delisle et al. 2012; Batygin \& Morbidelli 2013) and is in agreement with close-in planetary configurations detected by the Kepler mission among KOI's candidates. It should be noted that these works focus the attention on first order MMR (e.g. 2/1, 3/2, 4/3), while the case of co-orbital tidal evolution of exoplanets has not been still explored in the context of the general three-body problem. This is the main task of the present paper.

\subsection{Review of stable co-orbital configurations}

In the following, we describe the most important results concerning the equilibrium points of the co-orbital configurations in the general three-body problem. The reader is referred to Giuppone et al. (2012) for a summary of previous works.


Particularly, Hadjidemetriou et al. (2009) studied the motion close to a periodic orbit by computing the Poincar\'e map on the surfaces of sections. For this task, the symmetric families of stable and unstable motions were constructed and a previously unknown stable configuration was discovered, referred to as the Quasi-Satellite ($QS$) solutions.

Giuppone et al. (2010) constructed the families of periodic orbits in the vicinity of the 1/1 MMR using a semi-analytical method. The authors identified two separate regions of stability, symmetric and asymmetric,  defined by the behavior of the resonant angles $(\sigma, \Delta \varpi)\equiv (\lambda_2 \!-\!\lambda_1 , \varpi_2-\varpi_1)$, where $\lambda_i$ and $\varpi_i$ are mean longitudes and longitudes of pericenter of the planets, respectively. Summarizing:

\begin{itemize}
\item
$L_4$ and $L_5$ (asymmetric). Are the classical equilateral Lagrange solution associated to local maxima of the averaged Hamiltonian function. Independently on the mass ratio $m_2/m_1$, where $m_1,m_2$ are the masses of the planets, and eccentricities $e_1,e_2$, these solutions are always located at $(\sigma,\Delta\varpi) = (\pm 60^\circ,\pm 60^\circ)$. The size of the stable domains around these points decreases rapidly for increasing eccentricities and is practically negligible for $e_i > 0.7$. \\


\item
$AL_4$ and $AL_5$ (asymmetric). Anti-Lagrangian solutions which correspond to local minimum of the averaged Hamiltonian function. For low eccentricities, they are located at $(\sigma,\Delta\varpi) = (\pm 60^\circ,\mp 120^\circ)$. One anti-Lagrangian solution $AL_i$ is connected to the corresponding $L_i$ solution through the $\sigma$-family of periodic orbits in the averaged system (the solutions with zero-amplitudes of the $\sigma$--oscillation; for detail, see Giuppone et al. 2010). Contrary to the classical equilateral Lagrange solution, their locations on the plane $(\sigma,\Delta\varpi)$ depend on the planetary mass ratio and eccentricity values. Although their stability domains also shrink with increasing values of $e_i$, these solutions survive at eccentricities as high as $\sim 0.7$.\\

\item
Quasi-Satellite (symmetric). Are characterized by oscillations around a fixed point which is always located at $(\sigma,\Delta\varpi) = (0,180^{\circ})$, independently on the planetary mass ratio and eccentricities. In contrast with the $L_4$ and $L_5$ configurations, the domain of these orbits increases with increasing eccentricities and it fills a considerable portion of the phase space in the case of moderate to high eccentricities. \\

\end{itemize}

\subsection{Formation of exo-Trojans}\label{origin}
There is a vast literature trying to explain the formation of co-orbital planets and different mechanisms have been proposed. Here we only highlight some of the proposed scenarios: planetesimal and planet-planet scattering (Kortenkamp 2005; Morbidelli et al. 2008), direct collisional emplacement (Beaug\'e et al. 2007), in situ accretion (Chiang \& Lithwick 2005) and migration in multiple protoplanet systems (Cresswell \& Nelson 2006; Hadjidemetriou \& Voyatzis 2011; Giuppone et al. 2012).

Particularly, Giuppone et al. (2012) analyzed whether co-orbital systems may also be formed through the interaction of two planets with a density jump in the protoplanetary disk. The authors considered two planets, initially located farther than the 2/1 MMR (beyond 1 AU, where tidal interaction with the central star is almost negligible) and involved in the inward migration process, which is ended by the capture in the 1/1 MMR. Assuming an isothermal and not self-gravitating disk, the capture of massive planets into the resonance is obtained when the mass ratio is sufficiently close to unity and the surface density of the disk is sufficiently high. Multiple planetary systems with Earth-like planets produced co-orbital configuration after some scattering between them. The final outcome showed co-orbital configurations producing more easily solutions around $L_4$ and $L_5$, and less probably the $AL_4$ and $AL_5$ configurations. If the planets has large mass ratios, the smaller planet was either pushed inside the cavity or trapped in another mean-motion commensurability outside the density jump.

In view of the results regarding the formation of co-orbital configurations, we restrict our investigation to the case of equal mass planets, considering super-Earth and Saturn examples. We will investigate which is the subsequent evolution under tidal effects after the 1/1 resonance capture.

\section{Dynamics of the conservative problem}\label{maps}

\subsection{Phase space of the 1/1 MMR}\label{models}

The 1/1 MMR can be analyzed through a Hamiltonian formalism as it has been done for other mean motion resonances (see Michtchenko et al. 2006, 2008 ab). Basically, it requires a transformation to adequate resonant variables and a numerical averaging of the Hamiltonian with respect to short-period terms. The reader is referred to Giuppone et al. (2010) for a detailed description of our semi-analytical approach.

\begin{figure}
\begin{center}
\includegraphics[width=0.8\columnwidth,angle=0]{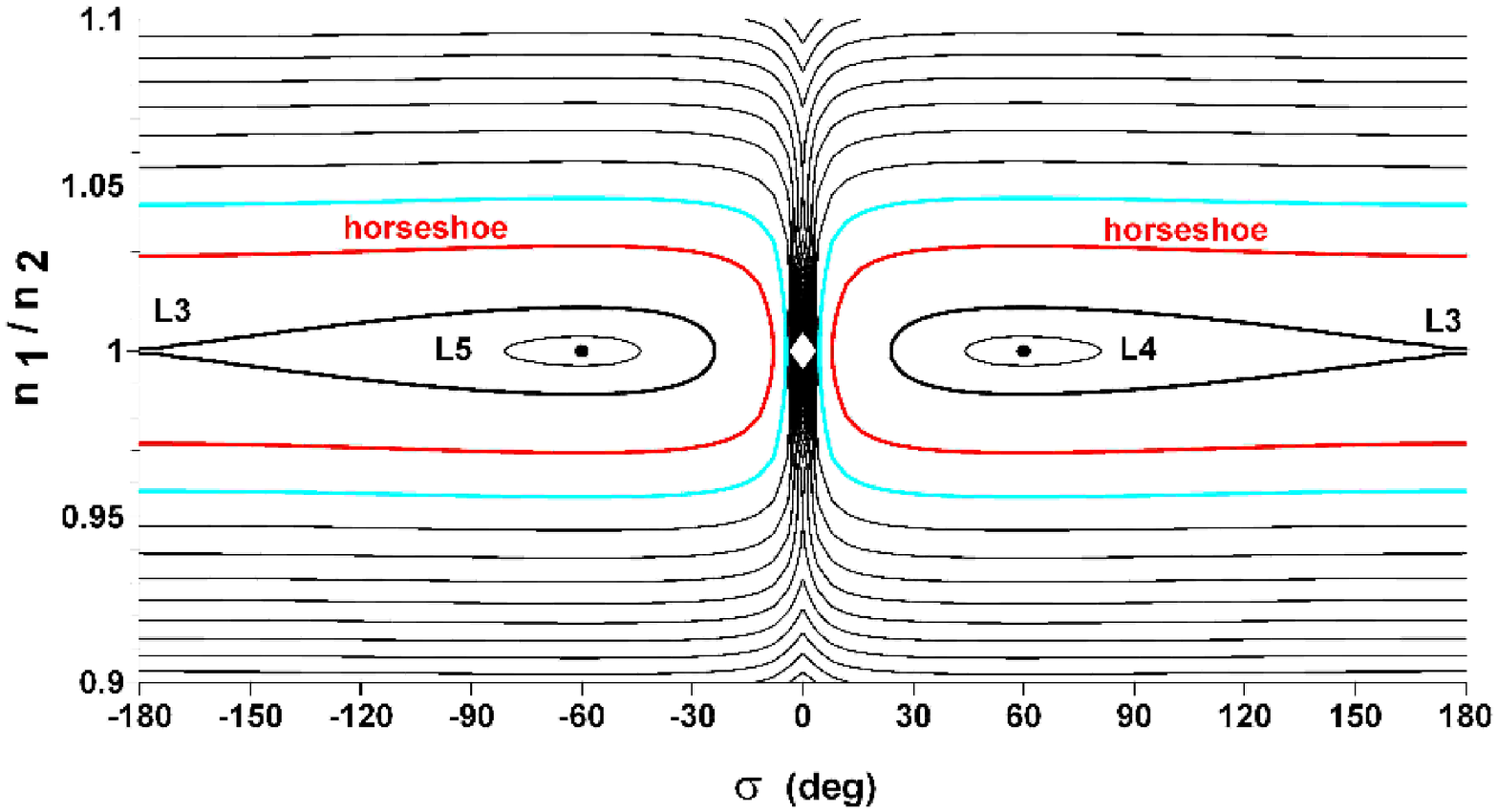}
\includegraphics[width=0.8\columnwidth,angle=0]{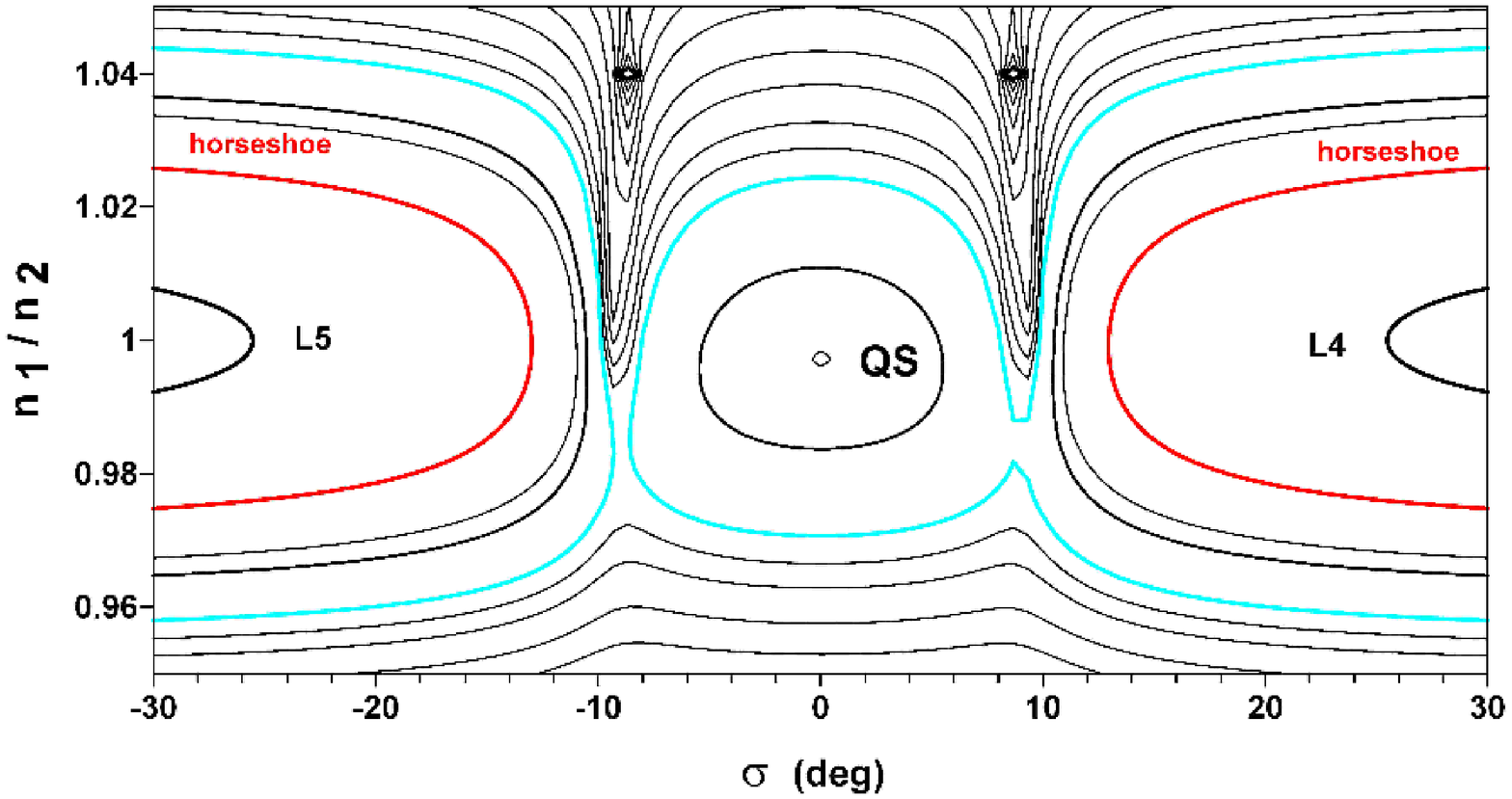}
\caption{\small Energy levels of the conservative Hamiltonian system composed of the Sun-like star and two co-planar planets of equal masses, $m_1=m_2=5m_{\oplus}$, on the ($\sigma,n_1/n_2$) plane. Top panel: initial values of the planet eccentricities are $e_1=e_2=0.0001$ and $\Delta\varpi=60^\circ$. Bottom panel: same plane (enlarged around the origin), obtained with $e_1=e_2=0.04$ and $\Delta\varpi=180^\circ$. Values of $\Delta\varpi$ and eccentricities were chosen for a better comparison with further results.}
\label{fase-st}
\end{center}
\end{figure}

Using a semi-analytical model we can visualize the phase space of the 1/1 MMR. For this task, we calculate level curves of the Hamiltonian which are plotted on the ($\sigma, n_1/n_2$) representative planes in Fig. \ref{fase-st}. Hereafter, we call $a_i,e_i,n_i$, for $i=1,2$, the semi-major axes, eccentricities and mean orbital motions of the planets, respectively. In the construction of the top panel we adopted $e_1=e_2=0.0001$ and $\Delta\varpi=60^\circ$, and, for the bottom panel, $e_1=e_2=0.04$ and $\Delta\varpi=180^\circ$; in each case, the planet masses were $m_1=m_2=5m_{\oplus}$, while the mass of the star was $m_0=1m_{\odot}$. The interpretation of the obtained portraits is simple, after some initial considerations. Since the averaged resonant system has two degrees of freedom, the phase space of the problem is four-dimensional and the intersection of one planetary path with the representative plane is generally given by four points belonging to the same energy level. However, for the chosen small values of the planet eccentricities, two degrees of freedom interact weakly each with other; in other words, they are nearly separable. This means that each energy level represents, with a good approximation, one nearly circular planetary path on the ($\sigma, n_1/n_2$) plane.

The stable stationary orbits of the planets in 1/1 MMR are represented by two fixed points in Fig. \ref{fase-st} (top panel); they correspond to the Lagrangian solutions $L_4$ and $L_5$ of the circular problem. Their locations in the phase space are given by $n_1/n_2=1$ and $\sigma=\pm 60^\circ$. The oscillations around $L_4$ and $L_5$ points are frequently referred to as tadpole orbits. Their domains are bounded by the separatrix shown by the thick black curves in Fig. \ref{fase-st}, which pass through the unstable saddle-like $L_3$ solution with coordinates
$n_1/n_2=1$ and $\sigma=\pm 180^\circ$. Outside the separatrix, there is a zone of horseshoe orbits (large amplitude oscillations of $\sigma$ around 180$^{\circ}$, encompassing both $L_4$ and $L_5$), which is extended up to the second  separatrix, which contains the Lagrangian saddle-like points $L_1$ and $L_2$ (not shown in Fig. \ref{fase-st}) located on the vertical axis at $\sigma=0$. Configurations leading to close encounters between the planets are represented with cyan curves. Such configurations are of special interest; indeed, we will show that the width of the regions of close approaches between both planets is correlated with the individual masses and that the short-term mutual interactions inside this region provoke imminent disruptions of the system. However, when the two planets are sufficiently close each to other, they form a specific dipole-like configuration known as quasi-satellite stable orbit. The domain of $QS$ is close to the origin of the plane and can be clearly seen in the amplified frame in Fig. \ref{fase-st} (bottom panel). Finally, the origin is a singular point which corresponds to collision between the planets.

It is worth noting that the orbits $L_4$ and $L_5$ correspond to the maximal values of the energy of the conservative 1/1 resonant system. This fact should be kept in mind when dissipative forces are introduced in the system. In this case, the variation of the energy will dictate the evolution of the resonant system: for slowly increasing energy, the system will converge to one of the $L_4$ and $L_5$ solutions, while, for slowly decreasing energy, its trajectory will be a spiral unwinding from $L_4$ (or $L_5$) solution. In the case of a dissipative evolution, the system starting nearly the $L_4$ equilibrium point will cross the domains of tadpole and horseshoe orbits, approaching the close encounters region, where it probably will be disrupted.

\subsection{Amplitude maps}\label{amp-maps}

\begin{figure*}
 \begin{center}
   \begin{tabular}{c c}
\hspace{-1cm}
\includegraphics*[width=7.5cm]{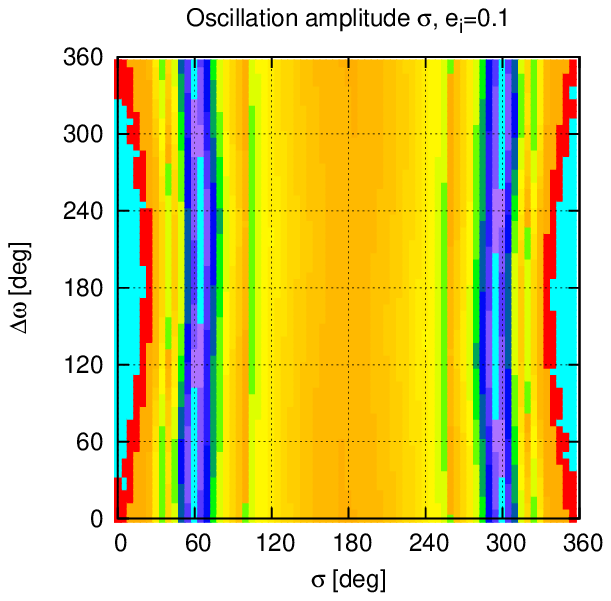} &
\hspace{-2.5cm}
\includegraphics*[width=7.5cm]{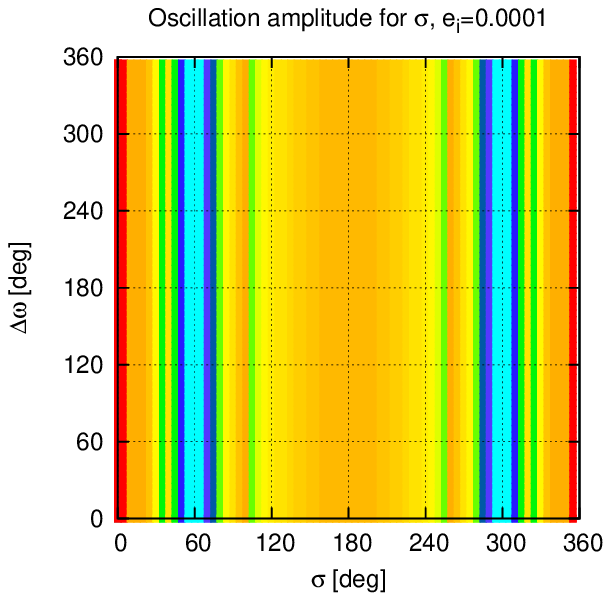}  \\
\hspace{-1cm}
\includegraphics*[width=7.5cm]{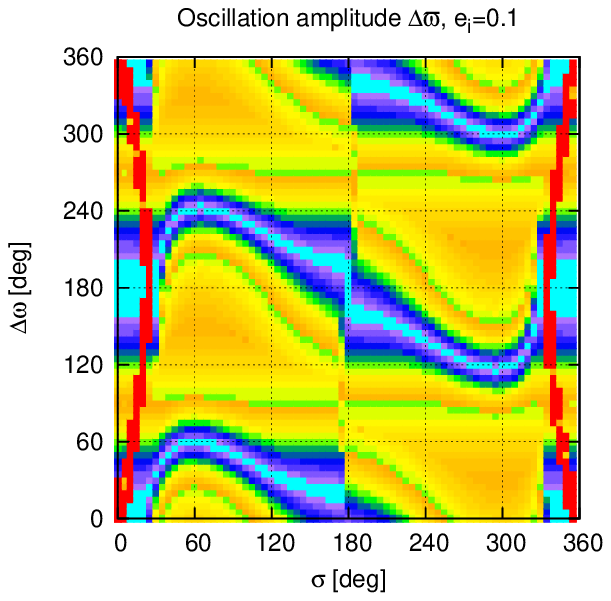} &
\hspace{-2.5cm}
\includegraphics*[width=7.5cm]{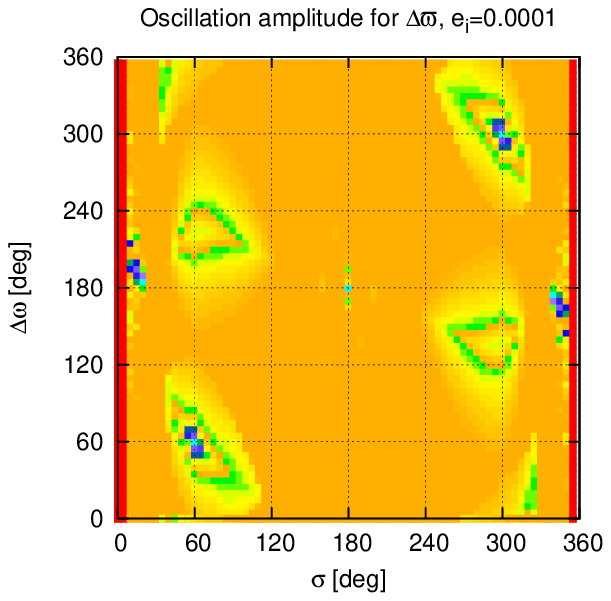}  \\
\multicolumn{2}{c}{\includegraphics*[width=6.cm]{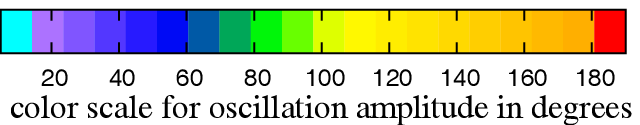}}
   \end{tabular}
 \caption{Amplitude maps for the resonant angle $\sigma$ (top panels) and the secular angle $\Delta \varpi$ (bottom panels) on the $(\sigma, \Delta \varpi)$--plane of initial conditions.  Results were obtained for a two super-Earth planets orbiting a Sun-like star with $a_1=a_2=0.04$ AU (orbital periods around 2.9 d); the initial eccentricities were chosen as $e_i$=0.1 (left column) and $e_i$=0.0001 (right column). Light-blue domains correspond to nearly zero amplitudes ($\sigma$- and $\Delta \varpi$-families of periodic orbits), darker regions indicate oscillation amplitudes smaller than $\sim 180^{\circ}$, while red color indicates collision orbits. For $e_i$=0.1 (left column), the phase space clearly shows $\sigma$--families as light-blue vertical strips on the top panel, and $\Delta \varpi$-families as horizontal strips on the bottom panel. The domains of unstable motion (in red) associated to close encounters between the planets surround the region of $QS$ located at $\sigma=0^{\circ}=360^{\circ}$. For nearly circular orbits, with 
$e_i$=0.0001 (right column), the domains of instability cover the region of $QS$ orbits. The location of the $\sigma$--families remains almost unaltered (top panel), in contrast with $\Delta \varpi$-familie, which almost disappear.}
\label{fig.dyn.earth}
 \end{center}
\end{figure*}

In this section, we present dynamical maps of the 1/1 MMR on the $(\sigma,\Delta\varpi)$ representative plane. For this task, we construct grids of initial conditions varying both $\sigma$ and $\Delta\varpi$ between zero and $360^{\circ}$. Each point in the grid was then numerically integrated over $5000$ years (roughly orbital 450 000 periods) using a Bulirsch-Stoer based N-body code. We calculated the amplitudes of oscillation of each angular variable. Initial conditions with zero amplitude in $\sigma$ correspond to {\it $\sigma$-family} periodic orbits of the co-orbital system, while solutions with zero amplitude in $\Delta \varpi$ correspond to periodic orbits of the {\it $\Delta \varpi$-family} (see Michtchenko et al. 2008ab). Stationary solutions of the averaged problem are the intersections of both families.

Fig. \ref{fig.dyn.earth} shows results obtained for a system composed of a Sun-like star and two hot-super-Earths with masses of $5m_{\oplus}$ and semi-major axes of 0.04 AU. The top and bottom frames show the oscillation amplitude of $\sigma$ and $\Delta \varpi$, respectively. Left and right frames were constructed with eccentricities $e_i=0.1$ and $e_i=0.0001$, respectively. The domains in light-blue color correspond to orbits with small amplitudes ($< 5^{\circ}$), thus indicating the location of the families of periodic orbits. The $\sigma$-families defined by nearly zero amplitudes of oscillation of the resonant angle $\sigma$ are located on the top frames. The $\Delta \varpi$-families defined by nearly zero amplitudes of oscillation of the secular angle $\Delta \varpi$ can be observed on the bottom panels. For example, at the initial condition with $e_i=0.1$ located at ($\sigma,\Delta\varpi$)=($120^\circ$,$40^\circ$), $\sigma$ oscillates with amplitude $100^\circ$, while $\Delta\varpi$ oscillates with very small amplitude around $40^\circ$. Darker regions correspond to increasing amplitudes and denote initial conditions with quasi-periodic motion. The unstable orbits (collisions) parameterized by amplitudes equal to $180.1^\circ$, are shown by red color.


We observe the domains around the Lagrangian equilateral solutions ($L_4/L_5$) at $(\sigma,\Delta \varpi)=(\pm 60^{\circ},\pm 60^{\circ})$, the anti-Lagrangian solutions ($AL_4/AL_5$) at $(\sigma,\Delta \varpi)\simeq(\pm 60^\circ,\mp 120^\circ)$, and the $QS$ solution $(\sigma,\Delta \varpi)=(0,180^{\circ})$.

For low eccentricities ($e_i=0.1$, left frames), we observe the asymmetric solutions connected with the $\sigma$-family of periodic orbits (vertical cyan strips) and their intersection with the horizontal {\it $\Delta \varpi$-family} (sinusoidal strip at bottom frame). The unstable orbits are very thin regions (in red) that separate the two regimes of motion (symmetric from asymmetric). 

For initially almost circularized orbits ($e_i=0.0001$, right frames), the $QS$ region disappears (vertical red strips at $\sigma=0$), meanwhile the low amplitude oscillations for the {\it $\sigma$-family} in top frame remains almost unaltered as vertical strips. In the bottom right frame, the {\it $\Delta \varpi$-family} shrinks into a concentrated region around the exact location of periodic families. Outside these small domains, $\Delta\varpi$ oscillates with high-amplitude ($>160^{\circ}$).

The results of the maps show that, during the process of tidal eccentricity damping, we expect an increase in the oscillation amplitude of the angles around the equilibrium points of the 1/1 MMR. This feature will be confirmed through analysis of the numerical simulations of the exact equations of motion. 

\begin{figure}
\begin{center}
\includegraphics[width=0.7\columnwidth,angle=0]{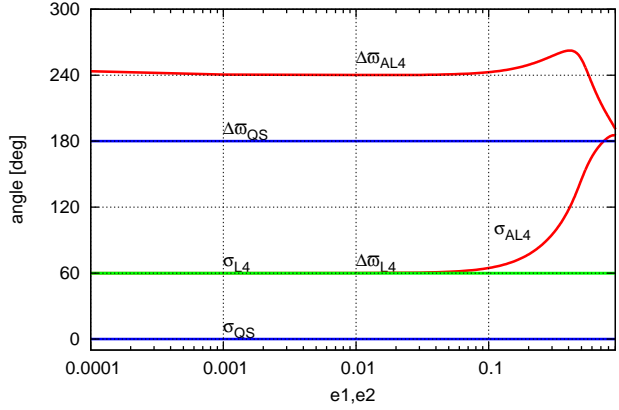}
\caption{Equilibrium values of $\sigma$ and $\Delta \varpi$ for the periodic families as function of the eccentricity for $m_2/m_1=1$.}
\label{fig.famil}
\end{center}
\end{figure}

Using the semi-analytical Hamiltonian model we constructed the families of periodic orbits following Giuppone et al. (2010). Fig. \ref{fig.famil} shows the dependence of the angular coordinates of the equilibrium points on the eccentricities from quasi-circular orbits to $e_i=0.9$. This figure will serve us as a guide in the choice of the initial conditions for the numerical simulations of the exact equations of motion (see next section). 

The results shown in this section are qualitatively the same for hot-Saturn planets.

\section{The equations of motion of the dissipative problem}\label{model}



We consider a system composed by a central star and two close-in co-orbital and rotating planets. Due to the short astrocentric distances, we suppose that both planets are deformed by the tides raised by the central star, which is also assumed distorted by the tides raised by the planets. The orbital planes of motion are supposed to be coincident with the reference plane (i.e, zero inclinations). In addition, we assume that the rotation axes are normal to the orbital planes (i.e, zero obliquities). For close-in planets, it is also convenient to consider the forces arising form general relativity, in addition to the mutual gravitational interaction and tidal forces. Remember, we call $m_0$, $m_1$ and $m_2$ the masses of the star and planets, respectively; whereas the radii are denoted by $R_0$, $R_1$ and $R_2$.

In a reference system centered in the star, where the positions and velocities of the planets are $\mathbf{r}_i$ and $\mathbf{v}_i$, with $i=1,2$, the equations of motion are given by

\begin{eqnarray}\label{mov1}
\ddot{\mathbf{r}}_1&=&-\frac{G(m_0+m_1)}{r_1^3}\mathbf{r}_1+Gm_2\Bigg{(}\frac{\mathbf{r}_2-\mathbf{r}_1}{|\mathbf{r}_2-\mathbf{r}_1|^3}-\frac{\mathbf{r}_2}{r_2^3}\Bigg{)}\nonumber\\
&&+\frac{(m_0+m_1)}{m_0m_1}(\mathbf{f}_1-\mathbf{f}_{01}+\mathbf{g}_1)+\frac{\mathbf{f}_2-\mathbf{f}_{02}+\mathbf{g}_2}{m_0},\\
\ddot{\mathbf{r}}_2&=&-\frac{G(m_0+m_2)}{r_2^3}\mathbf{r}_2+Gm_1\Bigg{(}\frac{\mathbf{r}_1-\mathbf{r}_2}{|\mathbf{r}_1-\mathbf{r}_2|^3}-\frac{\mathbf{r}_1}{r_1^3}\Bigg{)}+\nonumber\\
&&\frac{(m_0+m_2)}{m_0m_2}(\mathbf{f}_2-\mathbf{f}_{02}+\mathbf{g}_2)+\frac{\mathbf{f}_1-\mathbf{f}_{01}+\mathbf{g}_1}{m_0}.\label{mov2}
\end{eqnarray}
On one hand, $\mathbf{g}_i$ are the general relativity contributions acting on the planets, which for $i=1,2$ are given by

\begin{equation}\label{relativ}
\mathbf{g}_i=\frac{Gm_im_0}{c^2r_i^3}\left[\left(4\frac{Gm_0}{r_i}-\mathbf{v}_i^2\right)\mathbf{r}_i+4(\mathbf{r}_i\cdot\mathbf{v}_i)\mathbf{v}_i\right]
\end{equation}
where $\mathbf{v}_i=\dot{\mathbf{r}}_i$ and $c$ is the speed of light (see Beutler 2005). On the other hand, $\mathbf{f}_i$ are the tidal forces raised by the star acting on the masses $m_1$ and $m_2$ due to their deformations, respectively. We use the expression for tidal forces given by Mignard (1979):
\begin{equation}\label{mignard1-eq-pla}
\mathbf{f}_i=-3k_{i}\Delta t_i\frac{Gm_0^2R_i^5}{r_i^{10}}[2\mathbf{r}_i(\mathbf{r}_i\cdot\mathbf{v}_i)+r_i^2(\mathbf{r}_i\times\mathbf{\Omega}_i+\mathbf{v}_i)],
\end{equation}
where $\Omega_i$ is the rotation angular velocity of the $i$-planet, for $i=1,2$. It is worth noting that Mignard's force is given by a closed formula and, therefore, is valid for any value of eccentricity\footnote{Note however that, for a more accurate description, a large number of harmonics in the expansion of the tidal potential should be considered when the star-planet distance is small enough (see Taylor and Margot 2010).}. $k_{2i}$ is the second order Love number and $\Delta t_i$ is the time lag, which can be interpreted as a delay in the deformation of the tidally affected body due to its internal viscosity. The total tidal force on the star is $\mathbf{f}_0=\mathbf{f}_{01}+\mathbf{f}_{02}$, where $\mathbf{f}_{0i}$ are the individual tidal forces raised by each planet and for $i=1,2$ are given by

\begin{equation}\label{mignard1-eq-star}
-\mathbf{f}_{0i}=-3k_{0}\Delta t_0\frac{Gm_i^2R_0^5}{r_i^{10}}[2\mathbf{r}_i(\mathbf{r}_i\cdot\mathbf{v}_i)+r_i^2(\mathbf{r}_i\times\mathbf{\Omega}_0+\mathbf{v}_i)],
\end{equation}
where the subscript ``0" stand for star quantities. Note that are the forces $-\mathbf{f}_{0i}$ which act on the planets and must be considered in the equations of motion of the bodies.

The fact that $\Delta t\neq0$ introduces energy dissipation in the system, resulting in orbital and rotational evolution due to tidal torques. The tidal model here adopted is a classical linear approach (Darwin, 1880), since it is implicitly assumed that the resulting dissipation is proportional to the tidal frequencies. This tidal model is frequently referred to as a constant time-lag model and, despite that recent works have shown that it could not be appropriate for the study of terrestrial planets, is expected to yield to the approximately correct results. For a review of other tidal models, the reader is referred to Efroimsky \& Williams (2009) and Ferraz-Mello (2013).

\section{Numerical simulations}

In this section we show the result of the numerical integration of equations (\ref{mov1})--(\ref{mov2}) for some particular systems. As shown in Giuppone et al. (2010), large-scale planetary migration should favor the formation of co-orbital configurations more likely for nearly equal mass of the planets (see Sec. \ref{origin}). Hence, we restrict our investigation to the case $m_2/m_1=1$. We start with the case of two super-Earth planets with $m_1=m_2=5m_{\oplus}$ orbiting a Sun-like star with $m_0=m_{\odot}$ and $R_0=R_{\odot}$. The radii are such that the mean densities $\overline{\rho}$ satisfy $\overline{\rho}_1=\overline{\rho}_2=\overline{\rho}_{\oplus}$, so that $R_1=R_2=5^{1/3}R_{\oplus}$.

For sake of completeness we also explore the case of individual masses of $m_{Sat}$, where $m_{Sat}$ is the mass of Saturn, and the radii are computed as in the super-Earth case. 

We are going to investigate the final outcome of a system originally evolving in a co-orbital configuration when the tidal effect acts to change the orbital elements. Thus, we choose the initial values of the angles $(\sigma,\Delta\varpi)$ near to the equilibrium points. The initial semi-major axes and eccentricities are $a_1=a_2=0.04$ AU (orbital periods of $\simeq2.9$ d) and $e_1=e_2=0.1$. The initial values of rotation periods are $16.7$ h for both planets, however, they are not important because the rotation rapidly encounters its stationary value which only depends on the eccentricity within the adopted tidal model (Hut, 1981). We set 19.4 d as the initial value of the stars' rotation period, noting that the stationary value in this case is reached in a much larger timescale than for the planets. In addition, we need to know the value of the moment of inertia around the rotation axis, which is given by $\xi mR^2$, where $\xi$ is the structure constant. For super-Earths, we adopt Earth values for $\xi$, namely, $\xi=0.33$; whereas for hot-Saturn and a Sun-like star we set $\xi=0.21$ and $\xi=0.07$, respectively\footnote{http://nssdc.gsfc.nasa.gov/planetary/factsheet/}.   

The linear model enable us to relate the time delay $\Delta t$ with the quality factor $Q$ through $1/Q=n\Delta t$, where $n$ is the mean orbital motion (see Correia et al. 2012). However, the quality factor is poorly constrained even for Solar System bodies, although some previous works have brought valuable information. For instance, for the solid Earth we have $Q=370$ (Ray et al. 1996), and values of $Q/k_2=0.9\times10^{5}$, $Q>1.8\times10^4$ and $Q/k_2=4.5\times10^{4}$ were estimated for Jupiter, Saturn and Neptune, respectively (Lainey et al. 2009; Meyer \& Wisdom 2007; Zhang \& Hamilton 2008). More recent investigations (Ferraz-Mello 2013) suggest that the hot super-Earth CoRoT-7 b has $Q=49$, whereas for a hot Jupiter of 2--3 $m_{Jup}$ in a 5-d orbit, $Q\sim4.2\times10^5$.

In this work, we adopt $Q=20$, $k_2=0.3$ for hot super-Earths and $Q=1\times10^4$, $k_2=0.34$ for hot-Saturn planets, whereas for Sun-like stars we adopt $Q=1\times10^6$, $k_2=0.34$. The corresponding values of $k_2\Delta t$, where $\Delta t$ is obtained through $1/Q=n\Delta t$, are $(k_2\Delta t)_{super-Earth}=600$ sec, $(k_2\Delta t)_{Saturn}=1.37$ sec and $(k_2\Delta t)_{Sun}=0.0137$ sec.

We chose initial values of the angles within 4 degrees around the exact equilibrium solutions (see Fig. \ref{fig.famil} for the locations of the exact solutions and Table 1 for initial values chosen). The symmetric and asymmetric periodic solutions for mass ratio close to unity are such that $e_1=e_2$ (and also $a_1=a_2$, see Hadjidemetriou et al. 2009; Giuppone et al. 2010). Hence, we start our simulations with equal eccentricities, since we are assuming that the system is evolving under the 1/1 MMR. However, we note that the global results would not be affected by either the initial values of eccentricities or amplitudes of the angles, as long the corresponding configuration is nearby to the exact 1/1 MMR.  

For our stability criterion, we use the critical distance given by $d=\kappa(R_1+R_2)$. Hence, we assume that orbital instability would occur whenever the instantaneous mutual distance is equal or smaller than $d$, that is, for $|\mathbf{r}_2-\mathbf{r}_1|\leq d$. We note that $\kappa=1$ implies in the physical collision between the planets\footnote{The expression ``collision" in this work can also be used in association with instabilities of the planetary system (i.e, close encounters resulting in ejections or hyperbolic orbits, etc), thus, it is not restricted to impact between the planets. However, close encounters with $\kappa$ between 1 and 2, may result in tidal disruption of one or both planets
rather than ejection into hyperbolic orbit}. In this work, we use $\kappa=1$ and $\kappa=2$, however, we anticipate that the final results are not sensitive (in the sense of collision timescale) to the choice of one of the above values of $\kappa$.

\subsection{Results} 
   
\subsubsection{Hot super-Earths}  

\begin{table}\label{tab1}
\begin{center}
\begin{tabular}{c|c|c|c|c|c}
 &$m_2/m_1$ & $\sigma$ (deg) & $\Delta\varpi$ (deg) & $e_1$ & $e_2$   \\
\hline
 QS    & 1 &    4   &   184  &  0.1 & 0.1    \\
 $L_4$   & 1 &   64   &    64  &  0.1 & 0.1 \\
 $L_5$   & 1 &   296   &   296  &  0.1 & 0.1 \\
 $AL_4$  & 1 &   65   &   245 & 0.1 & 0.1 \\
 $AL_5$  & 1 &   295   &   115 & 0.1 & 0.1 \\
\end{tabular} 
\caption{Arbitrary initial conditions near the stable periodic solutions in the $(\sigma, \Delta \varpi)$ plane, calculated with the semi-analytical method (see Giuppone et al. 2010). All conditions have $a_1=a_2=0.04$ AU. The same initial co-orbital configurations were chosen for both hot super-Earth and hot-Saturn cases.}
\end{center}
\end{table} 

We start with the case in which the individual planets are represented by a $5m_{\oplus}$ super-Earth. 

\begin{figure}
\begin{center}
\includegraphics[width=0.358\columnwidth,angle=270]{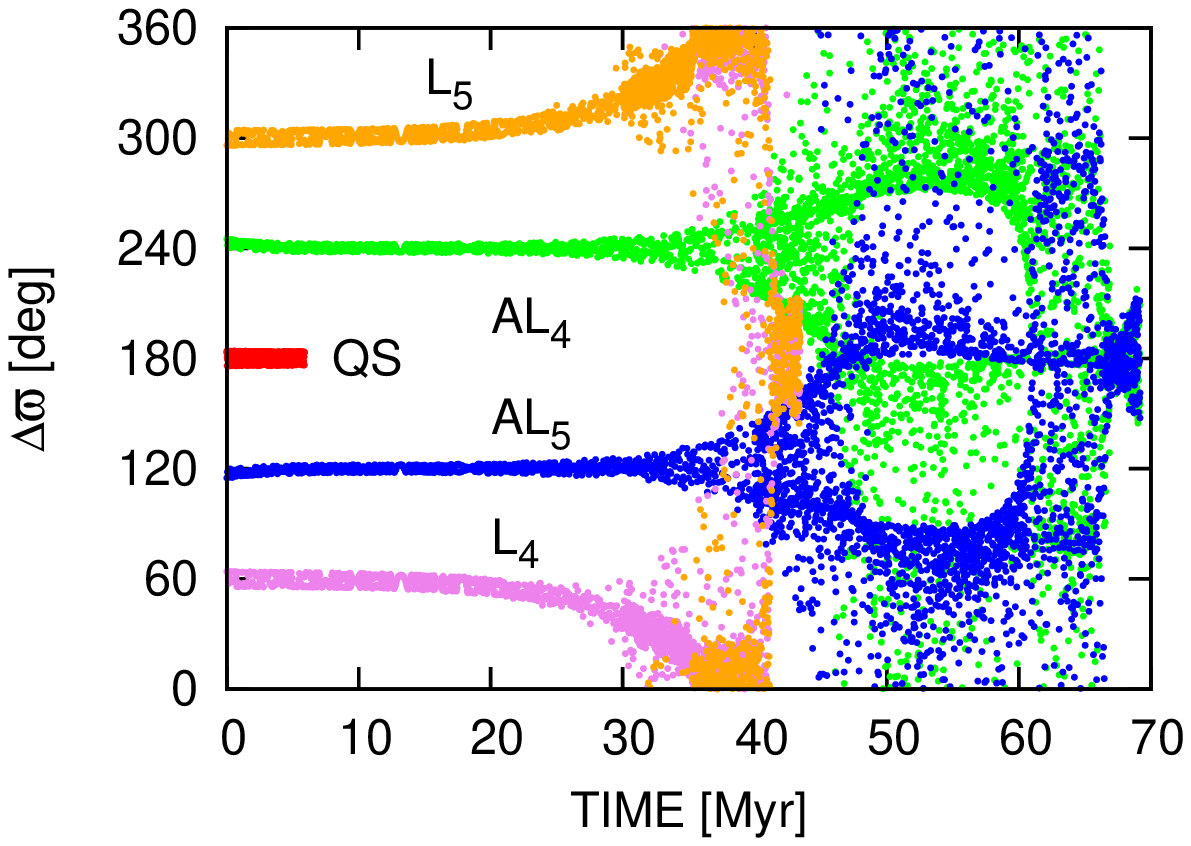}
\hspace{-0.5cm}
\includegraphics[width=0.358\columnwidth,angle=270]{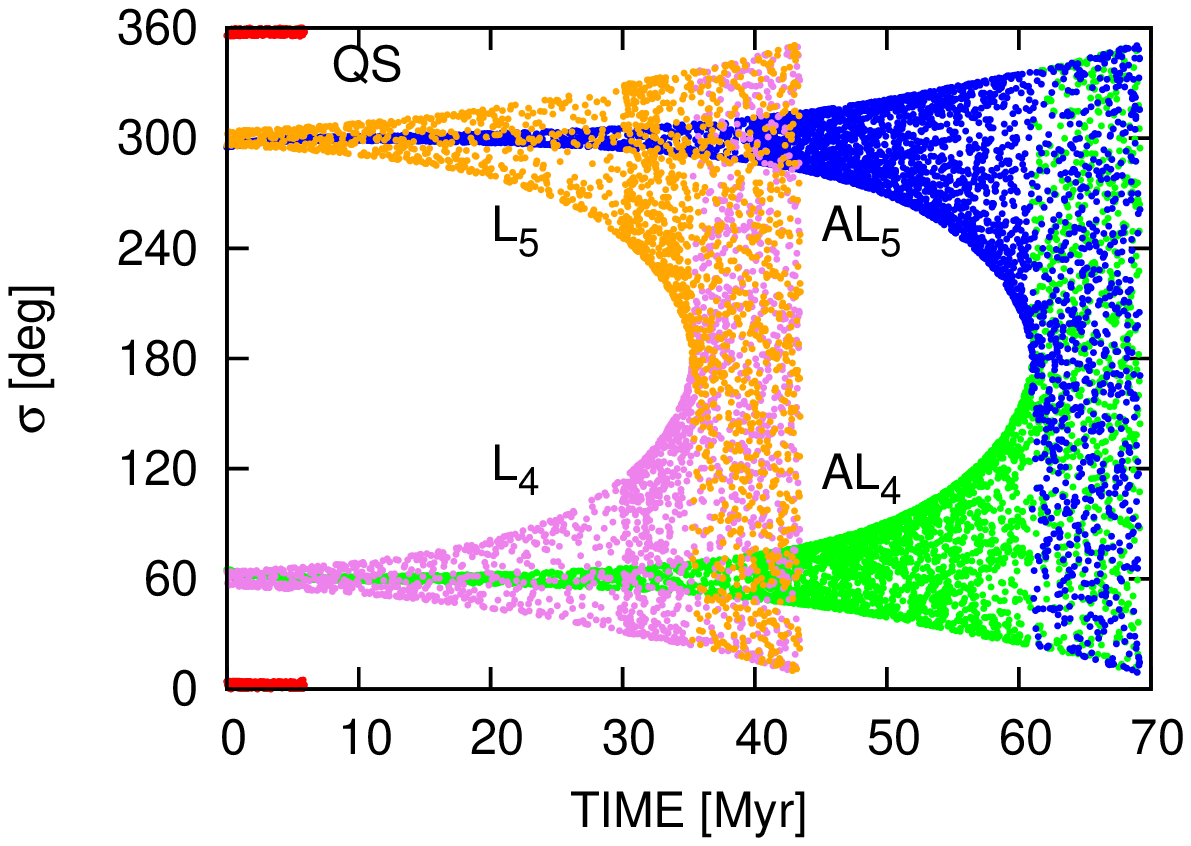}
\caption{\small Time variation of $\Delta\varpi$ and $\sigma$ (in degrees) for the symmetric ($QS$) and asymmetric ($L_4/L_5$ and $AL_4/AL_5$) co-orbital configurations, for starting values displayed in Table 1. Several regimes of libration/oscillation/circulation are present along the evolution (see text for detailed discussion). The planets ultimately collide on timescales of tens of millions of years, except for the $QS$ case in which the system destabilizes at approximately 6 Myr.}
\label{angles-st}
\end{center}
\end{figure}

\begin{figure}
\begin{center}
\includegraphics[width=0.5\columnwidth,angle=270]{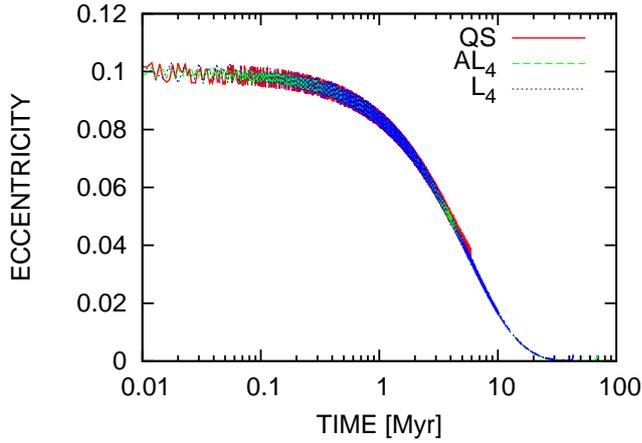}
\caption{\small Time variation of eccentricities for the system with two super-Earths planets. For $L_4$ and $AL_4$ co-orbital configurations, the circularization is obtained at the end of the simulations, whereas for the $QS$ case, the final values are close to 0.038. In all cases, the evolution follows close to the stationary solution $e_1=e_2$ (the values of the eccentricities are superimposed in the figure).}
\label{ecces-st}
\end{center}
\end{figure}

Figs. \ref{angles-st} shows the time variation of the angles $(\sigma,\Delta\varpi)$ around the symmetric and asymmetric equilibrium points for $\kappa=1$, whereas the evolution of the eccentricities is shown in Fig. \ref{ecces-st}. \textit{A collision between the planets is the final outcome of all simulations}. 

For $AL_4/AL_5$ configurations, the angle $\Delta\varpi$ oscillates with small amplitudes up to around 35 Myr. However, the amplitudes start to increase and, close to 45 Myr, a new regime of oscillation appears, in which the angles seem to avoid the regions centered near 240$^{\circ}$ and 120$^{\circ}$. Note that these positions are the corresponding equilibrium points of $\Delta\varpi$ for $AL_4/AL_5$ solutions in the circular case. Moreover, between 60\,--\,65 Myr, a circulation of $\Delta\varpi$ appears and, finally, an oscillation around 180$^{\circ}$ with increasing amplitude occurs before the collision between the planets.   

The evolutions of $\Delta\varpi$ along $L_4/L_5$ during the first 20 Myr follow close to the stationary values (60$^{\circ}$/300$^{\circ}$). After that, $\Delta\varpi$ deviates and a time interval of oscillation and circulation occurs, with the angle taking many values in the whole interval between 0$^{\circ}$\,--\,360$^{\circ}$ but privileging the oscillation around 0$^{\circ}$. Finally, $\Delta\varpi$ oscillates around 180$^{\circ}$ before collision.


Regarding to the resonant angle $\sigma$, we note in Fig. \ref{angles-st} a libration with increasing amplitude around the equilibrium points. When the regime of oscillation/circulation of the angle $\Delta\varpi$ appears, $\sigma$ librates around 180$^{\circ}$ with very high amplitude until the end of the simulation (for all asymmetric configurations). We will return to this discussion later in this section.  

The solution close to $QS$ is the first to become unstable, around 6 Myr of evolution. In this case, the amplitudes of the angles are close to 4$^{\circ}$ and $e_1\simeq e_2=0.038$ before destabilization. Several runs for the $QS$ case were carried out with different values of initial amplitudes (even zero) and the collision was the final outcome in all cases in very similar timescales ($<10$ Myr). Moreover, we tested a run without tides and the system becomes stable at least up to 60 Myr (with constant amplitudes of the angles).     

The evolution of eccentricities follows with $e_1\simeq e_2$ (and thus superimposed in Fig. \ref{ecces-st}), according to the stationary solutions for equal mass planets. The final result of the tidal evolution is an almost doubly orbital circularization except in the $QS$ case, in which the oscillation amplitude of eccentricities are larger than for other cases.

\begin{figure}
\begin{center}
\includegraphics[width=0.35\columnwidth,angle=270]{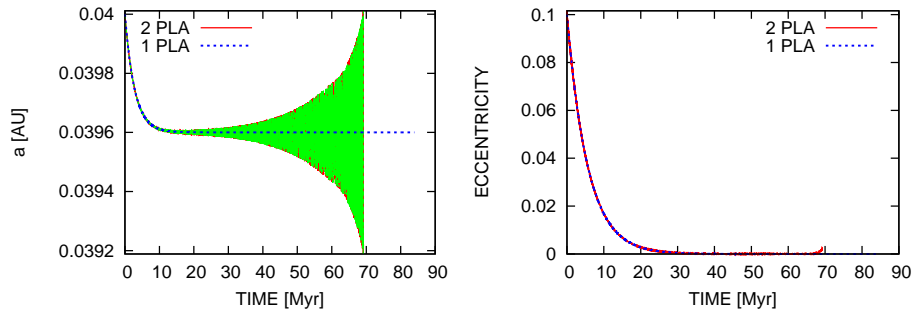}
\caption{\small Time variation of semi-major axes and eccentricities along the $AL_4$ solution (solid curves) and the comparison with the case of a single-planet (dashed curves), for the two super-Earths planet system. Both results are in good agreement, indicating that the evolution around the 1/1 trapping leads the planets to follow independent paths dictated by the tidal effect.}
\label{semi-ecce-st}
\end{center}
\end{figure}

If we suppose that the time variation of the eccentricities follows an exponential law (see right panel in Fig. \ref{semi-ecce-st}), we can estimate the timescale for tidal damping as $\tau_e=\dot{e}/e$. Using a classical averaged expression for $\dot{e}$ obtained from linear tidal models (see Eq. (3.5) in Rodr\'iguez \& Ferraz-Mello 2010) we obtain $\tau_e\simeq6$ Myr. This value gives an estimation of the e-folding of the eccentricity damping. Therefore, the timescales for destabilization shown in Fig. \ref{angles-st} correspond roughly to $10\tau_e$.

Fig. \ref{semi-ecce-st} shows the time variation of semi-major axes and eccentricities corresponding to the evolution along $AL_4$ (solid curves), noting that the condition of the equilibrium solution is followed ($a_1\simeq a_2$). In addition, we also plot the same elements when only one planet is present in the system (dashed curves). It is interesting to note that both results are in good agreement. The explanation can be found in the fact that, since the system follows an equilibrium configuration from the beginning provided by the resonant trapping, the tidal evolution acts independently to damp the orbital elements\footnote{Pauwels (1992) mentioned that when two satellites are locked in mean-motion resonance, the tidal effect will cause the orbits to expand independently.}. Moreover, the final value of semi-major axis in the case of a single planet can be found through $a_{fin}=a_{ini}\,\textrm{exp}(e_{fin}^2-e_{ini}^2)$, where the subscripts ``$ini$'' and ``$fin$'' stand for initial and final values, respectively (see Rodr\'iguez et al. 2011b). Replacing numerical values, we obtain $a_{fin}=0.0396$ AU, in good agreement with the numerical simulations. The above analytical calculation only considers the effect of planetary tides (tides on the planets), indicating that, in view of the agreement, the contribution of stellar tides (tides on the star) can be safely neglected\footnote{Indeed, the amplitude of stellar tides are proportional to $m_i/m_0$ (see Rodr\'iguez and Ferraz-Mello 2010).}.   


\begin{figure}
\begin{center}
\includegraphics[width=0.5\columnwidth,angle=270]{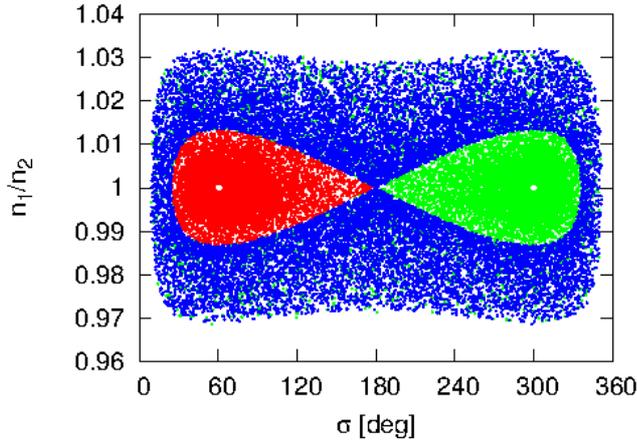}
\caption{\small The evolution in the plane $(\sigma,n_1/n_2)$ for the two super-Earth planets system around the $L_4/L_5$ equilibrium points. The structure of the resonant motion is identified through the domains of libration of the angle $\sigma$, disconnected by the corresponding separatrix. Similar results can be obtained for the $AL_4/AL_5$ equilibrium points.}
\label{semi-sigma-st}
\end{center}
\end{figure}

Fig. \ref{semi-sigma-st} shows the result of the numerical simulations for $L_4/L_5$ in the plane $(\sigma,n_1/n_2)$. Here, it is easy to identify the librations regimes of the resonant angle $\sigma$ within the domain of the 1/1 MMR. The initial small libration amplitude around $(60^{\circ},1)$ and $(300^{\circ},1)$ (red and green points) increases as the orbit circularizes due to the tidal evolution. When the amplitude is large enough, the motion occurs around $\sigma=180^{\circ}$ (blue points) and encompasses both equilateral Lagrangian points (see also Fig. \ref{angles-st}, right panel). In analogy with the restricted problem, these types of motion correspond to tadpole and horseshoe co-orbital configurations. In Fig. \ref{semi-sigma-st}, we clearly see that the tadpole and horseshoe oscillation regimes are disconnected by the separatrix of the 1/1 resonant motion.

We should note that the condition $\sigma\sim0$ is necessary for the collision to occur. However, we do not observe in Fig. \ref{semi-sigma-st} a separatrix disconnecting the regimes of libration and circulation of $\sigma$, and thus the angle seems never takes the zero value. To overcome this situation, we performed a new short numerical simulation starting very close to the configuration in which the collision occurs. We will illustrate these results in the next section in the application for hot-Saturn planets, but the same can be applied for super-Earths.

It is interesting to compare the numerical results with those predicted by the semi-analytical model (Sec. \ref{maps}). Indeed, the curve which separates the domains of tadpole and horseshoe motions (black curve on top panel of Fig. \ref{fase-st}) agrees with the separatrix appearing in Fig. \ref{semi-sigma-st}. The oscillation amplitude of $n_1/n_2$ in both cases is approximately 0.12, while $\sigma$ takes values as small as $25^{\circ}$ (for $L_4$), until the horseshoe domain arises.

The numerical solution for the $QS$ configuration indicates that $n_1/n_2\simeq1.02$ and $\sigma\simeq5^{\circ}$ before destabilization. We thus note an excellent agreement of the above result through an inspection of Fig. \ref{fase-st} (bottom panel), where the point $(\sigma,n_1/n_2)=(5^{\circ},1.02)$ belongs to the separatrix of the $QS$ co-orbital configuration (note that Fig. \ref{fase-st} for $QS$ was constructed taking $e_1=e_2=0.04$ for better comparison with the numerical results, in which the eccentricities are close to 0.038 just before collision).

\subsubsection{Hot Saturn}

The numerical exploration of the system with two Saturn planets indicates that, as in the previous system, \textit{collision between the planets results in all simulations for $\kappa=1$ and $\kappa=2$}. Moreover, all initial co-orbital configurations destabilize in timescales between 55-80 Myr, except the one around $QS$ in which the collision occurs near 430 Kyr.   

\begin{figure}
\begin{center}
\includegraphics[width=0.55\columnwidth,angle=270]{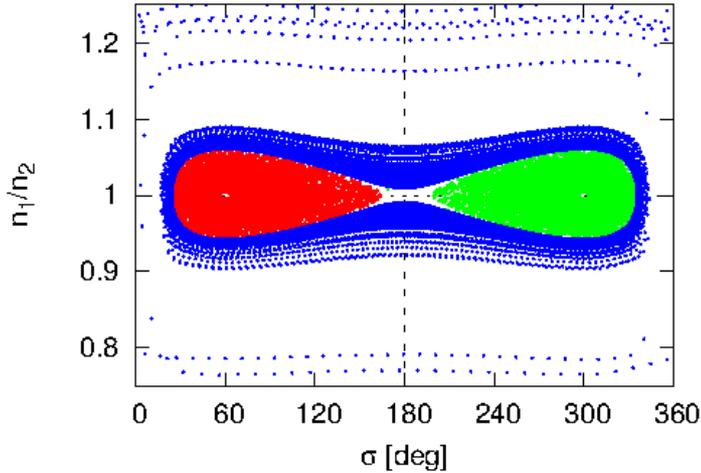}
\caption{\small The analogous to Fig. \ref{semi-sigma-st} for the system with two hot-Saturn planets. We can identify the domains of libration around the Lagrangian equilibrium points (tadpole and horseshoe) and also the circulation of $\sigma$. The collision takes place when $\sigma$ becomes very close to zero.}
\label{semi-sigma-sat-full}
\end{center}
\end{figure}

In analogy to Fig. \ref{semi-sigma-st}, Fig. \ref{semi-sigma-sat-full} displays the results of the numerical simulations for $\kappa=1$, showing the libration of $\sigma$ around the equilibrium Lagrangian points (tadpole and horseshoe orbits). In this case, the motion in the horseshoe domain is quite unstable due to the strong mutual interaction between the planets, and the system remains in that libration regime only 166 yr before collision (in the case of the two super-Earth system, the horseshoe regime lasted around 8 Myr). 

In addition to these types of libration, we see the domain in which the angle $\sigma$ circulates (also in blue points). In the circulation regime, in which the orbits are no longer locked in the 1/1 MMR, $\sigma$ takes the zero value several times, favoring the orbital conditions for the collision between the planets.



\section{Conclusions and discussion}\label{conclusions}

We investigated the motion of a two-planet system which evolves under combined effects of mutual and tidal interactions in the vicinity of a 1/1 mean motion resonance (i.e, co-orbital configuration). We considered systems of equal mass planets ($m_2/m_1=1$) corresponding to super-Earth ($5m_{\oplus}$) and Saturn planets ($\simeq95m_{\oplus}$). Numerical simulations of the exact equations of motion indicate that the collision between the planets is the final outcome of the tidal evolution for all initial co-orbital configurations tested in this work.  

We started our simulations considering initial conditions near the five stationary configurations of the co-orbital motion (namely, $L_4/L_5$, $AL_4/AL_5$ and $QS$). As tides continually damp the eccentricities, the initially small oscillation amplitudes of the angles $(\sigma,\Delta\varpi)$ around the equilibrium points increases and, ultimately, the instability occurs.  

We identified several libration regimes of the resonant angle $\sigma$, including the tadpole and horseshoe co-orbital configurations, in which the motion occurs around $\sigma=60^{\circ}/300^{\circ}$ and $\sigma=180^{\circ}$, respectively. In addition, the motion around the anti-Lagrangian equilibrium points ($AL_4/AL_5$), located close to $(\sigma,\Delta\varpi)=(60^{\circ}/300^{\circ},\\240^{\circ}/120^{\circ})$ for $m_2/m_1=1$ and small eccentricities, remains stable over a timescale about 1.5 larger than for the Lagrangian points ($L_4/L_5$) for both types of systems. Moreover, the stability of the quasi-satellite ($QS$) co-orbital configuration ($\sigma=0^{\circ},\Delta\varpi=180^{\circ}$) is restricted to short timescales ($<10$ Myr and $<1$ Myr, for super-Earth and Saturn individual masses, respectively).

The interpretation of the tidal evolution of the 1/1 resonant planet pair and its ultimate disruption is in the following. We know that the $L_4$ and $L_5$ stationary solutions correspond to global maxima of the conservative Hamiltonian of the 1/1 MMR, at least for small and moderate eccentricities. This is a consequence of the fact that a mean-motion resonance, acting as a protection mechanism, implies the maximal possible mutual distance between two planets at conjunction, where their closest encounters occur (Michtchenko et al. 2008b).  When the tidal interactions are introduced in the system, the energy of the system is dissipated through tidal heating of the planets. In this way, starting near one of the $L_4$ and $L_5$ points, the system evolving under dissipation will suffer the increase the oscillation amplitudes of the angles around the exact positions of the equilibrium points. It will cross the domains of the tadpole and horseshoe orbits, as described in Sec. \ref{maps}. Thus, the system will ultimately tend toward a collisional route. 


Therefore, the results presented in this work suggest that the tidal evolution of close-in planetary systems in the vicinity of the 1/1 MMR is globally unstable, constraining the possible detection of hot exo-Trojans (particularly among KOI's candidates). However, we have to stress that the origin of such close-in systems is unknown. Indeed, as recently shown in Giuppone et al. (2012), the formation of co-orbital configurations at distances of 1 AU is possible, however, the question on how did they reach close-in configurations is still under discussion. Also, it should be kept in mind that, for planets more distant from the central star, we expect a timescale for tidal evolution much longer than the age of the systems, and thus the possibility of radial velocity detections cannot be ruled out. 



As a final considerations, some limitations of our model should be highlighted. First, the applicability for rocky planets of the adopted linear tidal model has been recently questioned in some works (see Efroimsky \& Williams 2009; Ferraz-Mello 2013 and references therein). However, we speculate that the adoption of a different tidal model should change the timescales of the tidal evolution but the global result would not be affected.  

Second, we neglected the polar and permanent equatorial deformations of the bodies (i.e, $J_2$ , $C_{22}$ and higher orders terms). Recently, Rodr\'iguez et al. (2012) have shown that the contribution of $C_{22}$ can lead to temporary captures in spin-orbit resonances of rocky planets (see also Correia \& Rodr\'iguez (2013) and Callegari \& Rodr\'iguez (2013) for a calculation of $J_2$ and $C_{22}$ for a specific resonant rotation trapping). Moreover, the orbital decay and the eccentricity damping are larger whenever the planet rotation is trapped in a resonant motion. Hence, the tidal model and the consideration of the equatorial permanent deformation should be taken into account in further investigations.

Finally, following the suggestion of Dr. M. Efroimsky (private communication), we performed an additional simulation including an indirect tidal term which is described in the following. The tidal effect on the star due to one of the planets creates a (dissipative) torque which affects the orbital and rotational evolution of the tidal raising planet. The other planet will also ``feel'' this tidal effect because the star deformation crates a potential in an arbitrary point in the space. The effect of this force may be important for planets trapped in mean-motion resonances.
In the co-orbital case, we have a ``frozen'' configuration in which the relative positions of the bodies are located in the vertices of an equilateral triangle of variable size. We only consider the static tidal component (i.e., the instantaneous response, independent on $\Delta t_0$, see Ferraz- Mello et al. 2008), because it is orders of magnitude larger than its dissipative counterpart. The result of the numerical simulation for the two Saturn planets system have shown that the inclusion of these terms only slightly modifies the timescales of the planet evolution, delaying the instability in approximately 10\% for a $L_4/L_5$ starting configuration.

\begin{acknowledgements}
We acknowledge to M. Efroimsky for comments and suggestions. A.R and T.A.M acknowledge the support of this project by FAPESP (2009/16900-5) and CNPq (Brazil). C.A.G acknowledges the support by the Argentinian Research Council, CONICET. This work has made use of the computing facilities of the Laboratory of Astroinformatics (IAG/USP, NAT/Unicsul), whose purchase was made possible by the Brazilian agency FAPESP (grant 2009/54006-4) and the INCT-A. Some of the computations were performed on the Blafis cluster at the Aveiro University. We also acknowledge the two anonymous reviewers for their valuable comments and suggestions.  

\end{acknowledgements}

\end{document}